# Modelling the Political Context in Requirements Engineering. The System is made for Man, not Man for the System

Rana Siadati, Paul Wernick and Vito Veneziano
School of Computer Science
University of Hertfordshire
Hatfield, Hertfordshire (UK)
r.siadati@herts.ac.uk

**Abstract -** This paper describes the authors' point of view on reaching a stage at which it is necessary to understand customer organisations better to identify their problems and how to address them. To resolve this issue we need a mechanism to capture and model how that organisation actually operates by mapping organisations against the system to be developed, by including power and politics in their "too human" and even emotional dimension. We then describe here a notation by which to recognise and document power, politics and the emotional aspects of software requirements-related decision-making in customer organisations. We conclude by outlining that our suggested graphical notation would maybe not solve the problem: but even if it just raises awareness, this would make us closer to solving the problem. Given the sensibility of the political issue, it is assumed that the generated diagram using the above mentioned notations is only for the requirements engineer and his/her team, thus remaining strictly private.

## I. Background and context

Despite some progress, software practitioners are still some distance from fully mastering the art of eliciting, analysing and validating requirements in such a way that all parties would find it satisfactory.

Politics and the power have been identified as crucial components of requirements engineering (RE) and argue that the role it plays, especially when RE is applied to the software industry, needs to be given greater attention than is currently the case. The intention is to improve the understanding of professionals and academics of the present state of that role, and to face (if not to drive) any future evolution of the discipline by providing sounder conceptual and interpretative tools and models than are currently available. We contend that over-simplified views and considerations of such aspects have become predominant in how we train requirements engineers: such views may well have contributed to a selective blindness for power dynamics and how they do not always propagate linearly, from top to bottom, but rather follow more complex patterns.

We also feel that the adoption across the field of notations and technical language(s) from engineering (e.g., organigrams and UML) with limited ability to express, for example, ambiguity, to represent complex phenomena like organisations, can result in models that only capture a static, structural view, as if complex, changing webs of personal relationships in an organisation can be the object of just another engineering blueprint. This has in our opinion led to an implicit decision to ignore or abstract away how organisations become permeated by political relationships in a fluid, dynamic and sometimes unpredictable way.

If software requirements engineers and analysts are to be able to take advantage of any insight they could gain into politics and power relations within organisations – for example, who de facto decides which priorities come first when a conflict arises between two departments in the same company – a more appropriate (but unfortunately less easy) approach would have demanded, and will continue to demand, more effort to be made to understand the relevant political dynamics and their source(s) behind the official organisational structure. However, politics as actually occurring in an organisation should be seen as a fourth, more fundamental dimension informing the results of the above three tasks, one which sometimes subverts the 'official' power structures as might be documented as above. The informal chains of power must be understood as well as the formal when, for example, trading off requirements. We have now started research into identifying elements of a workable notation for documenting political and power relationships within a typical RE project, and through which the necessary negotiation processes might be contextualised and understood.

## II. Introduction

As software engineers our goal is to produce high-quality software. To achieve this we must ensure that we understand customers' needs not fail to meet their requirements/expectations, and they must know what to expect. As Alexander (2005) says, "The System is made for Man, not Man for the System."

Software Engineering emerged as a concept 50 years ago at NATO Garmisch conference in 1968 (Naur and Randell, 1969). Although many issues were raised and discussed, and since then we have tried very hard to find solution for our development problem, we are still failing to produce the highest quality software that our customers deserve. Nuseibeh et al. (2000) note that "The primary measure of success of a software system is the degree to which it meets the purpose for which it was intended. Broadly speaking, software systems requirements engineering (RE) is the process of discovering that purpose, by identifying stakeholders and their needs, and documenting these in a form that is amenable to analysis, communication, and subsequent implementation."

In many cases our failure is due to a failure in identifying crucial requirements. Hofmann and Lehner (2001) state that "deficient requirements [is] the single biggest cause of software project failure", adding that "getting requirements right might be the single most important and difficult part of a software project."

To gain success in the future we need to learn from the past and also to identify the factors that played a key role in this success. If we call the time from 1950s to 1968 the first generation of computerisation, the battle was to make systems work with the limited hardware available.

A second generation tried to produce more reliable systems better fitted to the organisation. Many languages, methods and techniques have been introduced in the last 50 years in attempts to achieve this; Randell (2018) states that "I am reluctant to accept that it justifies anywhere near 8945 languages, and the very large number of different methods and techniques that have been created." However, problems still exist.

We believe that we have now reached a stage at which it is necessary to understand customer organisations better to identify their problems and how to address them. It has not been uncommon for us to expect customers to incline towards our (software developers') ways of thinking about their organisations and systems, whilst ignoring the internal dynamics of customer organisations. We suggest that software requirements can only be agreed on all sides if we understand the way that customer's organisation (ref Ian's paper about not one customer) makes decisions.

We have also previously stated (2017a) our belief that "since 1995, both practitioners and academics have not done enough to address non-technical issues, and/or that some crucial factors that might possibly improve RE practice have not yet been effectively addressed." We also raised there the question of why has the focus of RE developments been mostly on the technical component.

To resolve this issue we need a mechanism to capture and model how that organisation actually operates. We have previously argued (2007b) that argue that "this issue could be successfully addressed and resolved if, when we map organisations against the system to be developed, we include power and politics in their "too human" and even emotional dimension."

We describe here a notation which recognises and documents power, politics and the emotional aspects of software requirements-related decision-making in customer organisations. We suggest that "a simple way to do so is to use emoji pictograms: most of them are part of a universal language, which requirements engineers could easily adopt and exploit to assess and produce models that include an extra layer of "political" information in existing organograms, without the need to actually introduce a radically new notation.

## III. Technical and non-technical factors

Geethalakshmi and Shanmugam (2008), along with many other software engineers, developers and/or authors, point out that the success and failure of any software development project depends not only on technical factors, but on other non-technical factors/components. Non-technical factors have the same amount of influence, if not more, than the technical factors on the success or failure of software development projects.

Despite the concern of Hull et al., (2002) that "the most common reasons for project failures are not technical" for many years new techniques (addressing the technical component instead) have been suggested, although Fricker et al (2015) states that "many of these techniques did not become industrial because they were not practicable or ineffective when used in real-world projects." Even when new techniques and tools have been introduced to address technical problems, we are still at times failing to deliver successful projects or to even increase the success rate. We believe that, as the project development life cycle starts from system requirements specification, the effect of non-technical factors, whether obvious or hidden, should be considered at this level to prevent failure at a later stage.

## IV. Organisational politics

We have identified organisational politics as one non-technical factor which has existed for as long as organisations themselves. What could be beneficial is to recognise or even highlight important entities such as organisational politics and demonstrate it in our modelling which could lead to express the software design. In Organisational Behavior (2010), Brandon and Seldman (2004) and Hochwarter et al. (2000) were cited for stating that "organizational politics are informal, unofficial, intentional/unintentional and sometimes behind-the-scenes efforts to sell ideas, influence an organization, increase power, or achieve other targeted objectives".

However, we observe that little work has been done to date to assist requirements engineers navigate organisational politics to gain acceptance for sets of systems requirements. Milne and Maiden (2012) write that 'although notable work has been undertaken on the importance of social factors in RE, there has been relatively little direct consideration of the influence of power and politics'.

We believe that modelling the actual power relationships in an organisation, as against those identified from a traditional organogram, will help the requirement engineer identify those influences which not necessary always comes from the person above. It is possible to have a scenario in which the influencer not to be the powerful person in the formal hierarchy, when influences go beyond the formal to include the informal influences both within and outside formal structures. Knowledge of these situations will assist a requirements engineer to understand how to achieve a solution which will be acceptable to those most able to influence requirement decisions. In particular, we feel that is it vital for a requirements engineer to have this information available when they take a job over from a colleague who is already aware of, and may be taking into account, the need to convince informal as well as formal power-holders.

Betts (2011) believes that "IT professionals have to deal with corporate politics - in fact, they need to embrace it". We've all heard "techies" say: "Leave me out of the politics. I just want to implement the technology." But that's not a recipe for success. As the book puts it: "Where there's technology, there's change, and where there's change, and there will be people who perceive themselves a winner or loser. That's where politics begin."

## IV. Modelling politics alongside design of the future system

Adopting the right modelling technique (or techniques) is also another challenging task and usually some sort of abstract language or diagrammatic representation is used in modelling techniques. Non-diagrammatic requirements modelling techniques are widely spread. However, Beatty and Chen (2012) claim that "visual requirements models are one of the most effective ways to identify software requirements. They help the analyst to ensure that all stakeholders—including subject matter experts, business stakeholders, executives, and technical teams—understand the proposed solution.

Visualization keeps stakeholders interested and engaged, which is key to finding gaps in the requirements. Most importantly, visualization creates a picture of the solution that helps stakeholders understand what the solution will and will not deliver".

## VI. What modelling notations should be used to show the politics in a software engineering project?

It is common experience that designing the future system usually implies modelling it by using different notations.

There are lots of different notations or approaches for modelling stakeholders' and system requirements, such as UML. Modelling notations can be used to demonstrate and assist the requirements engineers' understanding of the problem.

Beatty and Chen (2012) state that "to make a requirements process 'fly', the first step is to understand that there is more than one kind of requirements model. A shopping list of requirements is invaluable in a contract, but on its own, it's desperately difficult to check for correctness and completeness, and it doesn't offer any suggestions on how to discover requirements, either. Different requirements models are needed to assist with discovering, checking, and analysing the requirements. The 'shopping list' is an output, not the one-and-only input."

However, we still need a simple notation to document both informal and formal power relationships.

Our suggested graphical notations for modelling requirements would maybe not solve the problem:

but even if it just raises awareness, this would make us closer to solving the problem.

Given the sensibility of the political issue, it is assumed that the generated diagram using the above mentioned notations is only for the requirements engineer and his/her team, thus remaining strictly private.

**Table 1 - A notation for graphical modelling of politics in requirements engineering**

| Name | Notation | Comment |
| --- | --- | --- |
| Entity | 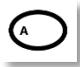 | Identification will be presented by circled entity (with name, title or other identification) |
| Formal Power Relationship | 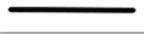 | Single lines will be used as connectors and represents power within the organisation. More lines show more power. |
| Power Direction | 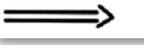 | Lines with arrow shows the direction of influence/power which can only be either one sided. |
| Power to Block | 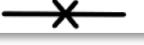 | Lines with a cross in the middle represent power to block which means that person is not reachable but he/she might be reachable through another person. For instance, a manager will not allow a requirements engineer to talk to his employees as he might find out about the influence/power within the organisation. |
| Note: Status of stakeholders will be shown by emoji faces which can be defined as some internal or external entity that interacts with the system. The purpose of using emoji faces including facial expression. | | |
| Happy Stakeholder | 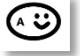 | 'Happy' stakeholder (satisfied with current system requirements) |
| Sad Stakeholder | 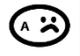 | 'Sad' stakeholder (dissatisfied with current system requirements) |
| Neutral Stakeholder | 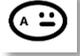 | Neutral stakeholder |
| Informal Relationship/influence | 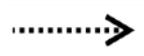 | Informal relationship/influence |

## VII. Discussion

An important aspect of requirements engineering is that people may change their view of the benefits or disadvantages of one or more system requirements as a result of instructions from more senior members of the organisation, or for reasons or influences outside the scope of the formal organisational structure. We show below a couple of examples of how the notation would reflect changing attitudes to a proposed system:

**Fig 1. Outcome of Informal relationship/influence: change of views/status**

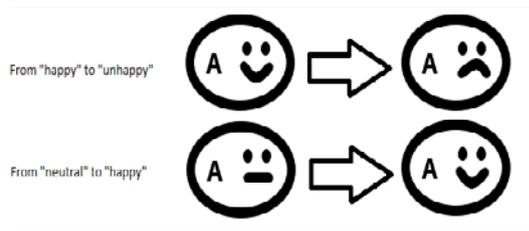

We set out here a how we would document a small organisation, and how informal power structures might affect the acceptability of a set of system requirement.

Figure 2 below show a typical organisational structure (or an organigram), such as might be found when documenting a hierarchical organisation. This is what is immediately visible from aspects such as organisation charts, job titles, formal roles, size of offices and quality of office carpets and so on.

**Fig. 2. Traditional organigram.**

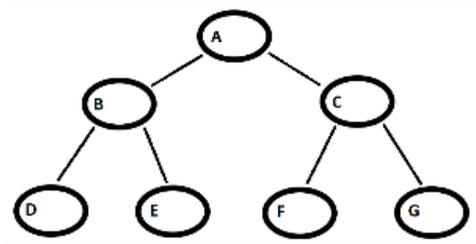

It is generally the case that those nearer to the top of this pyramid have more power than those nearer to the bottom. To show our assumption of the direction of power we add arrowheads to the lines as shown in Figure 3.

**Fig 3. Directions of power (the usual assumption behind the organigram)**

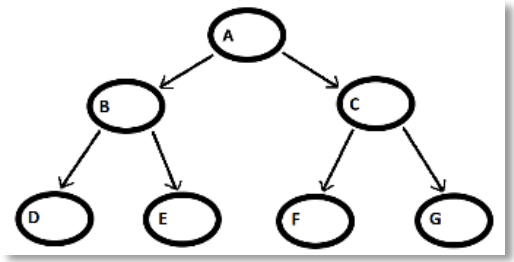

This diagram represents who reports to whom in a company, and who can formally direct subordinates. However, it does not capture the degree to which a senior member of the structure can direct their subordinates – their relative power. To represent this, we add more liens to the more powerful relationships, as shown in Figure 4, in which A has more power over B than they have over C, and C has more power over F than they have over G, and even more than A has over B. This may be due to the job roles, some of which are more directed than others; compare, for example, the relative power positions of a finance manager over a finance clerk whose work is routine and follows a set of rules laid down by their seniors, with the power of a design director over product designers whose creativity and independence may be valued rather than discouraged.

**Fig. 4. Organigram showing relative power (with multiple arrows)**

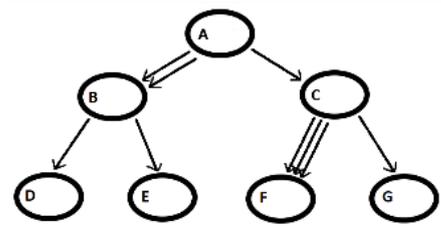

The next important influence on the work of the requirements engineer is that of the opinion of the user and decision-maker community on the system requirements as currently stated. Our notation therefore needs to show how happy or unhappy each participant is. This can be added to our diagram by the use of Emoji faces to show the status of stakeholders.

Fig 5 adds information on the opinions of each person in the hierarchy as to the system requirements. This examples shows that all are currently happy with these, with the exception of one low-level person (D). Ostensibly this might indicate that there will be no problems in obtaining support from people with sufficient authority to obtain agreement for the requirements.

**Fig. 5. Organigram showing relative power and status/views of each person.**

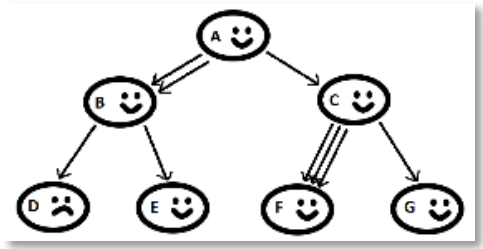

However, not all the relevant relationships and influences are reflected in the formal structures documented so far. For the requirements engineer to understand the complete political environment within which they are working is essential to show *informal* organisational and other relationships.

Figure 6 shows an example of an informal relationship between two entities which can change the whole situation. Here for some reason the apparently lower-level member of the hierarchy has an influence in decision making beyond that which would be expected from their apparent lack of formal power. This might be due for example, to a back channel within the organisation, or a personal relationship outside work between the two people, such as a common sport or private relationship. Documenting this is important for the requirements engineer to have a full understanding of how the influences on decision making operate; it may also sometimes show the importance of maintaining the privacy of this model from people who might not be aware of the relationship being documented.

In Figure 6, D has a relationship, whose specific nature is not specified here, with A. Whereas Figure 6 suggests that D's unhappiness with two requirements can be ignored because of their comparatively low position in the formal hierarchy, an *informal* influence on A might cause A to change their mind to a greater or lesser extent, and thus cause A to act or decide in a manner which would be unexpected in the absence of this information.

Other internal or external people, roles or entities that interact with the system and will have influence over decisions on its requirements can also be represented in this notation using the dotted 'informal influence' lines shown below.

**Fig. 6. Informal influence of D on A**

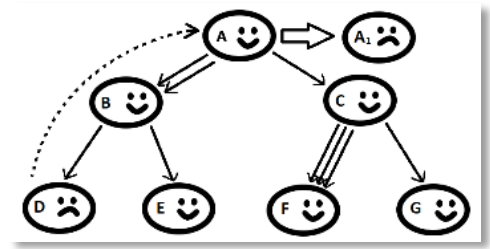

In the above figure, the informal relationship and power/influence of D over A has turned A from Happy to unhappy mode. Give that A has more control to B than they have over C, the mode of B may also be changed from happy to unhappy whilst C remains happy. This situation is reflected in Fig 8 below.

**Fig. 7: Changes of views/status after informal influence of D on A**

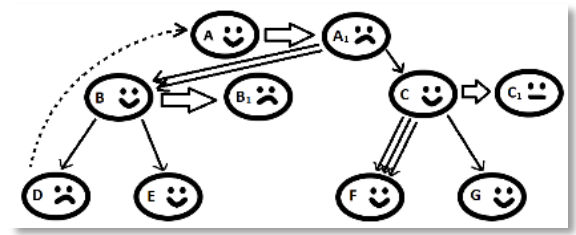

To summarise, we assume (and the model now reflects) that the overall decision-maker (A) was happy with the work of the requirements engineer but then becomes unhappy (A1) under the informal influence of D, which may trigger a need to change the requirements. Should this possibility have been known earlier, the requirements engineer might have talked and listened more to D who is an informal but significant influencer.

## VIII. Conclusions and future work

The suggested graphical notation for modelling the political context in RE would maybe not solve the problem: but even if it just raises awareness, this would make us closer to solving the problem. The suggested notations is to capture, at least partially, important features of any political relationship within organisations and enrich their modelling outcomes. As mentioned in Siadati R et al. (2017a), "at the moment, practitioners in RE can exercise their professional expertise by being "aware" of the political dimension, or by simply assuming the engineering process is politically neutral." We shall now adopt the Impact Evaluation methodology, based on the retrospective counterfactual analysis of what difference an intervention would have made in outcomes. We acknowledge this is an area worthy of further investigation and argue its outcomes could produce simple and yet effective tools, which practitioners can actually use in their daily activity.

The next step will be inviting experienced practitioners to use our approach to analyse their previous projects and consider whether this would have helped them in their work, particularly in identifying and resolving political and power-related related issues which they had to address. We have already identified a number of volunteers to help us in this work; if you would like to join us, please contact the corresponding author via email (r.siadati@herts.ac.uk).

In the meantime, we will be seeking more information and feedback from experts in both industry and academia.

**References**

1) Alexander, I. F. (2005), A Taxonomy of Stakeholders, Human Roles in System Development. International Journal of Technology and Human Interaction, Vol 1, 1, 2005, pages 23-59.
2) Beatty, J. and Chen, A. (2012), Visual models for software requirements. Pearson Education.
3) Betts, M. (2011), No, you can't avoid office politics in IT. Deal with it. Computerworld, Available: https://Www.Computerworld.Com/Article/2470396/It-Careers/No--You-Can-T-Avoid-Office-Politics-In-It--Deal-With-It-.Html (accessed 01/09/2018)
4) Brandon, R. and Seldman, M. (2004), Survival of the savvy: High-integrity political tactics for career and company success. Free Press, New York.
5) Fricker S.A., Grau R., Zwingli A. (2015), Requirements Engineering: Best Practice. In: Fricker S., Thümmler C., Gavras A. (eds), Requirements Engineering for Digital Health. Springer, Cham.
6) Geethalakshmi, S.N. and Shanmugam A. (2008), Success and Failure of Software Development: Practitioners' Perspective. Proceedings of the International Multi Conference of Engineers and Computer Scientists 2008, Vol I IMECS 2008, 19-21 March, 2008, Hong Kong.
7) Hochwarter, W. A., Witt, L. A., & Kacmar, K. M. (2000). Perceptions of organizational politics as a moderator of the relationship between conscientiousness and job performance. Journal of Applied Psychology, 85, 472–478.
8) Hofmann, H.F. and Lehner, F. (2001), Requirements Engineering as a Success Factor in Software Projects. IEEE Software, 18, 4, 58-66.
9) Hull E., Jackson K. and Dick J. (2002), Requirements Engineering. Springer, Cham.
10) Milne, A. and Maiden, N. (2012), Power and Politics in Requirements Engineering: A proposed Research Agenda. Requirements Engineering Journal, Springer, 17, 2, 83-98.
11) Naur P. and Randell B. (Eds.), 1969. Software Engineering: Report on a Conference Sponsored by the NATO Science Committee. Garmisch, Germany, 7-11 Oct. 1968.
12) Nuseibeh B. and S. Easterbrook (2000). Requirements Engineering: A Roadmap. ICSE '00 Proceedings of the Conference on The Future of Software Engineering, 35-46. ACM, New York.
13) Organizational Behavior (2010). Authors and publisher removed. ISBN: 978-1-946135-15-5. https://doi.org/10.24926/8668.1501
14) Randell, B. (2018), Fifty Years of Software Engineering or The View from Garmisch. ICSE, May 2018.
15) Siadati, R., Wernick, P., & Veneziano, V. (2017a). Politics for the Requirements Engineer. Submitted to IREB.
16) Siadati, R., Wernick, P., & Veneziano, V. (2017b). Modelling politics in requirements engineering: adding emoji to existing notations. arXiv:1703.06101.